\documentclass[10pt,conference]{IEEEtran}

\usepackage{graphicx}
\usepackage{textcomp}
\usepackage{xcolor}
\usepackage{xspace}
\usepackage{tcolorbox}
\usepackage{booktabs,multirow,stfloats}
\usepackage{fontawesome}
\newcommand{\approach}{{\sc STEAM}\xspace}
\def\BibTeX{{\rm B\kern-.05em{\sc i\kern-.025em b}\kern-.08em
    T\kern-.1667em\lower.7ex\hbox{E}\kern-.125emX}}

\begin{document}

\title{STEAM: \textbf{S}imulating the In\textbf{T}eractive B\textbf{E}havior of Progr\textbf{AM}mers for Automatic Bug Fixing}

\author{
\IEEEauthorblockN{Yuwei Zhang\IEEEauthorrefmark{1}\IEEEauthorrefmark{2}, Zhi Jin\IEEEauthorrefmark{3}, Ying Xing\IEEEauthorrefmark{4}, Ge Li\IEEEauthorrefmark{3}}
\IEEEauthorblockA{\IEEEauthorrefmark{1}
Institute of Software, Chinese Academy of Sciences, Beijing, China}
\IEEEauthorblockA{\IEEEauthorrefmark{2}
University of Chinese Academy of Sciences, Beijing, China}
\IEEEauthorblockA{\IEEEauthorrefmark{3}
Key Laboratory of High Confidence Software Technologies (Peking University), Ministry of Education, Beijing, China}
\IEEEauthorblockA{\IEEEauthorrefmark{4}
School of Artificial Intelligence, Beijing University of Posts and Telecommunications, Beijing, China \\
zhangyuwei@otcaix.iscas.ac.cn, zhijin@pku.edu.cn, xingying@bupt.edu.cn, lige@pku.edu.cn}
}

\maketitle

\begin{abstract}
Bug fixing holds significant importance in software development and maintenance. Recent research has made notable progress in exploring the potential of large language models (LLMs) for automatic bug fixing. However, existing studies often overlook the collaborative nature of bug resolution, treating it as a single-stage process. To overcome this limitation, we introduce a novel stage-wise framework named \approach in this paper. The objective of \approach is to simulate the interactive behavior of multiple programmers involved in various stages across the bug's life cycle. Taking inspiration from bug management practices, we decompose the bug fixing task into four distinct stages: bug reporting, bug diagnosis, patch generation, and patch verification. These stages are performed interactively by LLMs, aiming to imitate the collaborative abilities of programmers during the resolution of software bugs. By harnessing the collective contribution, \approach effectively enhances the bug-fixing capabilities of LLMs. We implement \approach by employing the powerful dialogue-based LLM---ChatGPT. Our evaluation on the widely adopted bug-fixing benchmark demonstrates that \approach has achieved a new state-of-the-art level of bug-fixing performance.
\end{abstract}

\section{Introduction}
\label{int}
Software systems, by virtue of their inherent complexity and inadequate testing, inevitably contain bugs that can lead to substantial losses, encompassing financial impacts and potential risks to human life \cite{wong2017familiar}. To expedite the resolution of software bugs, automatic bug fixing \cite{monperrus2023living} has been proposed as a means to mitigate the costs associated with software debugging. The primary objective of bug fixing is to efficiently correct software bugs while facilitating timely software maintenance. The advancements in deep learning (DL) have sparked an increasing interest in neural-based bug fixing approaches \cite{zhong2022neural,zhang2023survey}, which exploit the powerful representation capabilities of DL models to autonomously learn intricate bug-fixing patterns. Nonetheless, existing neural-based approaches \cite{lutellier2020coconut,jiang2021cure,zhu2021syntax,zhang2023neural} heavily rely on historical bug-fixing datasets acquired from open-source repositories for supervised training, which may limit their effectiveness to a specific set of bug-fixing patterns and hinder their generalization to unseen bug types \cite{xia2022less}.

In recent years, the rapid advancements in generative artificial intelligence (AI) have spurred researchers to utilize large language models (LLMs) for tackling various software engineering (SE) tasks \cite{ozkaya2023application}. LLMs undergo unsupervised training using billions of open-source text/code tokens to achieve comprehensive language modeling. The utilization of diverse data sources during LLM training enriches their cross-domain knowledge, thereby facilitating effective generalization for corresponding downstream tasks.

Acknowledging the limitations of prior neural-based approaches, recent research has commenced exploring the potential of LLMs for automatic bug fixing without the necessity of fine-tuning. At present, the application of LLMs to bug fixing \cite{xia2023automated,prenner2022openai,jiang2023impact,sobania2023analysis} involves devising prompts that can either consist of the buggy code alone or a combination of the buggy code and a few bug-fixing examples. The goal is for LLMs to learn from the provided prompts and generate appropriate patches for the given buggy code. However, current LLM-based studies predominantly focus on the handling of buggy code and approach bug fixing as an end-to-end manner. Intuitively, even for experienced programmers, generating correct patches for complex bugs solely based on code implementation remains a significant challenge. Programmers, by nature, tend to seek teamwork, involving interaction and collaboration, as a means of tackling intricate tasks in SE practices \cite{mcChesney2004communication,lindsjorn2016teamwork}.

\begin{figure}[htbp]
  \centering
  \includegraphics[width=0.48\textwidth]{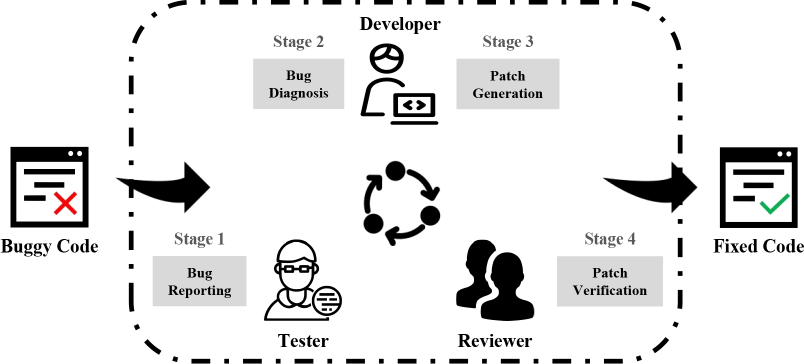}
  \caption{The Brief Structure of \approach.}
  \label{brief}
\end{figure}

More recently, researchers have demonstrated the remarkable capabilities of LLMs (e.g., ChatGPT \cite{openai2023chatgpt}) in generating helpful outcomes when tasks are disassembled into a set of modular units with precise queries \cite{wei2022chain,merow2023ai}. Bug fixing is a multifaceted task, wherein each discovered bug typically undergoes a specific and intricate process before it can be effectively resolved. However, existing bug fixing approaches based on LLMs generally treat the task as a single-stage process, neglecting the interactive and collaborative nature of programmers during the resolution of software bugs. To bridge the gap between the capabilities of LLMs and programmers in bug fixing, this paper introduces a stage-wise framework, referred to as \approach, which \textbf{S}imulating the in\textbf{T}eractive b\textbf{E}havior of multiple progr\textbf{AM}mers involved at various stages of the bug's life cycle. Drawing inspiration from bug management practices, we closely examine the programmers engaged in the bug management process and analyze the impact of their interactions on improving the efficiency of bug fixing. As depicted in Fig.\ref{brief}, \approach aims to imitate the collaborative problem-solving abilities exhibited by programmers (i.e., tester, developer, and reviewer) throughout the entire life cycle of a bug. Recognizing the significance of an efficient bug management process for successful bug fixing \cite{ohira2012impact}, we decompose the task into four distinct stages: bug reporting, bug diagnosis, patch generation, and patch verification. To be specific, effective bug resolution initially relies on the tester's comprehensive understanding of the bug, leading to the filing of a detailed bug report. This report provides essential information to the developer for resolving the bug successfully. Within the framework, the developer has a two-fold responsibility. Firstly, the developer engages in the diagnosis process by consulting historical bug corpus and conducting self-debugging. Secondly, the developer generates the candidate patch, guided by the information obtained in the previous stages. Since the correctness of the candidate patch generated on the first attempt cannot be guaranteed, the tester's involvement in \approach becomes crucial. The tester provides review feedback and collaborates with the developer throughout the workflow, playing a vital role in ensuring the correctness of the generated patch.

In summary, \approach breaks down the bug fixing task into smaller, manageable subtasks with the aim of improving the accuracy of automatic bug fixing through efficient bug management practices. Moreover, by involving multiple programmers, the proposed framework can harness diverse perspectives and feedback to facilitating the bug-fixing process, thereby mitigating misunderstandings and ensuring the quality of generated patches. Given the remarkable advancements in generative AI, LLMs have exhibited commendable performance across various SE tasks, opening avenues for inter-model interaction and collaboration. Therefore, our objective is to design system components that mimic the cognitive processes of programmers engaged in the proposed framework. Specifically, \approach employs three ChatGPT agents, each playing the role of a different type of programmer (i.e., tester, developer, and reviewer). Following system instructions, these agents simulate the corresponding programmer behaviors. \approach effectively aligns the collaborative abilities of the programmers by utilizing specific prompts, enabling an interactive bug-fixing process. In other words, the task of bug fixing is re-defined as a workflow, comprising simpler bug management stages where the outputs from earlier stages are used to construct the inputs for subsequent stages. The main contributions of this paper can be summarized as follows:
\begin{itemize}
  \item We present the first attempt at enhancing the capabilities of LLMs in automatic bug fixing by leveraging effective bug management practices. Our proposed alignment approach simulates the interactive behavior of programmers engaged in bug management, which enables LLMs to collaborate and generate correct patches for given bugs.
  \item We propose a stage-wise framework called \approach, consisting of three ChatGPT agents, each responsible for specific stages within the bug management process via system instructions and prompts.
  \item We conduct extensive experiments on publicly available bug-fixing benchmarks and thoroughly evaluate each component of the proposed framework. The experimental results demonstrate that \approach surpasses state-of-the-art baselines, highlighting its superior performance.
\end{itemize}

The remainder of this paper is organized as follows. We describe the related work in Section~\ref{rel}. Section~\ref{met} introduces in detail the proposed framework. We provide the experimental setup in Section~\ref{exp}. Section~\ref{res} shows the analyzing results of our research. We disclose the threats to the validity of our approach in Section~\ref{thr}. Section~\ref{con} draws conclusions and indicates directions for future work.

\section{Related Work}
\label{rel}

\begin{figure*}[t]
  \centering
  \includegraphics[width=\textwidth]{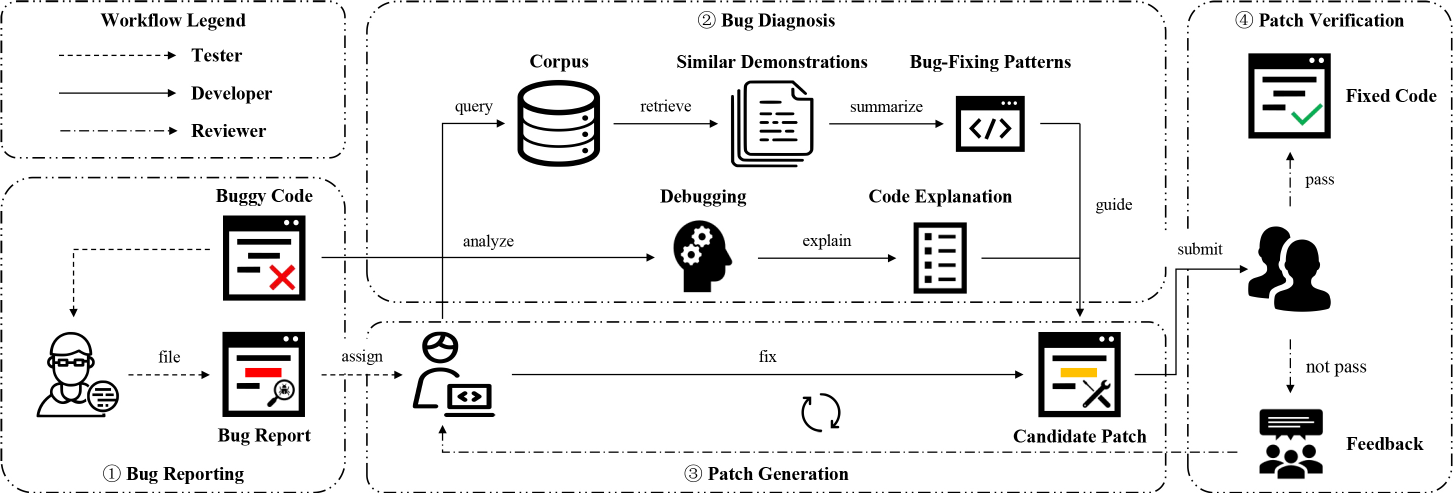}
  \caption{Overview of \approach.}
  \label{overview}
\end{figure*}

\subsection{Automatic Bug Fixing}

Over the last decade, automatic bug fixing has emerged as a promising research topic, garnering considerable attention from both the SE and AI communities. Traditional approaches \cite{monperrus2023living} can be broadly divided into two mainstream categories: search-based \cite{goues2012genprog,saha2019harnessing,ghanbari2019practical,liu2019tbar,koyuncu2020fixminer} and semantics-based \cite{mechtaev2016angelix,xuan2017nopol,le2017syntax,chen2017contract,afzal2021sosrepair}. With the rapid advancement of DL techniques, there has been a growing focus on neural-based approaches \cite{zhong2022neural,zhang2023survey}, which have shown remarkable potential for enhancing bug fixing performance. In contrast to traditional approaches, learning-based techniques possess the ability to automatically capture semantic relationships between parallel bug-fixing pairs. This capability allows for the generation of more effective and context-aware patch solutions. Nevertheless, candidate patches generated by neural models are typically not evaluated against the test suite or subjected to other automated verification strategies, as commonly done in traditional approaches. Consequently, they may encounter issues related to compilability. Most recently, researchers have delved into the feasibility of employing potent LLMs for automatic bug fixing. LLMs exhibit the capability to directly generate correct patches based on the surrounding context, obviating the necessity for fine-tuning. Despite the unprecedented outcomes achieved by LLM-based approaches \cite{xia2023automated,prenner2022openai,jiang2023impact,sobania2023analysis}, these techniques primarily concentrate on the buggy code and treat the bug fixing process as a single-stage task, disregarding the interactive and collaborative nature inherent in bug resolution. This paper introduces a stage-wise framework comprising multiple ChatGPT agents, each assigned to distinct stages within the bug management process using specific prompts. To the best of our knowledge, this is the first attempt to enhance the bug-fixing capabilities of LLMs through interactive simulation of programmer behavior.

\subsection{Large Language Model}

Recent developments in generative AI have led to a remarkable surge in performance and widespread adoption of LLMs \cite{xin2023survey}. LLMs undergo initial pre-training using a vast corpus that comprises both natural language text and source code. As LLMs are designed to be general and capable of acquiring knowledge from diverse domains, researchers can subsequently leverage LLMs for corresponding downstream tasks by providing tailored prompts or, optionally, a few demonstrations of the task being solved as input \cite{liu2023pretrain}. Among the LLMs, the GPT family \cite{radford2018improving,radford2019language,brown2020language,chen2021evaluating,openai2023chatgpt} by OpenAI stands out for its popularity and prowess. Additionally, many attempts have been made to reproduce similar open-source LLMs, such as CodeGPT \cite{lu2021codexglue}, CodeGen \cite{nijkamp2023codegen}, InCoder \cite{fried2023incoder}, LLaMA \cite{touvron2023llama}, GPT-NeoX \cite{black2022gptneox}, and others. Despite their robust performance, LLMs sometimes struggle to produce accurate answers when faced with complex tasks. In response, researchers have proposed the use of chain-of-thought (CoT) prompting \cite{wei2022chain} to enhance the reasoning capability of LLMs in natural language processing tasks. CoT involves a sequence of intermediate reasoning steps in natural language that culminate in the final output. In addition to traditional LLMs, more recently, researchers have proposed LLMs trained using reinforcement learning to better align with human preference. Examples of such models include InstructGPT \cite{ouyang2022training} and ChatGPT \cite{openai2023chatgpt}, which are initially initialized from a pre-trained model on autoregressive generation and then fine-tuned using reinforcement learning from human feedback (RLHF) \cite{ziegler2019finetuning}. This fine-tuning process using human preference has resulted in improved abilities of these LLMs to comprehend input prompts and follow instructions to perform complex tasks \cite{bang2023multitask}. Notably, ChatGPT has achieved state-of-the-art performance in various SE tasks \cite{dong2023self,sobania2023analysis}. The objective of this paper is to draw insights from effective bug management practices to enhance the capabilities of existing LLMs in the task of bug fixing. In particular, our experimental results demonstrate that such alignment enables powerful dialogue-based LLM---ChatGPT to interact and collaborate, significantly outperforming traditional LLMs.

\section{Methodology}
\label{met}

Aimed at overcoming the limitations mentioned in Section~\ref{int} concerning existing approaches, we present a novel framework denoted as \approach, which is a programmer-like behavior-simulation framework to empower LLMs in the task of bug fixing. In this section, we elaborate on the detailed design of our proposed framework. 

\subsection{Overview}

As illustrated in Fig.\ref{overview}, \approach consists of three ChatGPT agents (i.e., tester, developer, and reviewer), each responsible for specific stages (i.e., bug reporting, bug diagnosis, patch generation, and patch verification) within the bug management process. The functional profile of each stage is outlined as follows:

\begin{enumerate}
  \item \textbf{Bug Reporting:} During the initial phase, the tester discovers the bug within the source code and proceeds to file a detailed report elucidating the nature of the bug. In practice, bug reports play a crucial role in bug fixing as they provide the developer with essential information of the discovered bug. These specific details greatly assist the developer in resolving the bug \cite{zimmermann2010what,zou2020practitioners}. To simulate the tester's behavior, \approach is designed to generate an initial bug report that outlines the underlying cause of the buggy code, which is then assigned to the developer for further handling.
  \item \textbf{Bug Diagnosis:} Upon receiving a bug report from the tester, the developer commences the diagnose process by utilizing the available information provided in the report. Practically, when assigned a newly discovered bug, the developer first consults historical bug corpora to extract bug-fixing patterns that shed light on the causes and resolutions of similar issues. This mining process aids in acquiring valuable knowledge pertaining to the reasons behind bug occurrences and the corresponding fixes \cite{osman2014mining,tan2021crossfix}. Furthermore, the developer engages in debugging practices, wherein they meticulously analyze the source code line-by-line and articulate their findings in natural language. This self-guided approach enhances the efficiency of bug fixing without relying on external expert guidance \cite{chen2023teaching,paul2023automated}. To mimic the developer's diagnosis behavior, \approach initiates by retrieving relevant bug-fixing demonstrations for pattern summarization. Then, \approach employs rubber duck debugging techniques to provide explanations for the source code. These two forms of guidance serve to aid the developer in generating correct patches.
  \item \textbf{Patch Generation:} Once the root cause of a bug has been identified, the developer embarks on the process of creating a patch to fix the discovered bug. In previous studies, LLMs are instructed to generate patches directly based on the given buggy code. However, bug fixing is an intricate task that poses challenges in generating correct patches from scratch. Guided by the bug diagnosis process, the developer leverages bug-fixing patterns and code explanations as the prompt to produce a candidate patch, which is subsequently submitted to the reviewer for further verification. 
  \item \textbf{Patch Verification:} Generating correct patches in a single attempt can be particularly challenging for complex software bugs. As a result, the review process for patches assumes considerable significance as a pivotal activity within software peer review \cite{rigby2012contemporary,wang2015comparative}. When presented with a candidate patch generated by the developer, the reviewer needs to assess its effectiveness in resolving the discovered bug. In cases where the reviewer does not pass the candidate patch, indicating that the discovered bug has not been successfully fixed, the developer is required to make further modifications. Once the reviewer passes the candidate patch, the discovered bug is considered resolved. To simulate the behavior of reviewer, \approach first determines the suitability of the candidate patch as a correct solution for the buggy code. If the candidate patch does not meet the required criteria, \approach offers the review feedback to the developer for patch regeneration. This patch verification process might iterate several times between the reviewer and the developer until the candidate patch is deemed correct.
\end{enumerate}

Through the utilization of good bug management practices, \approach can enhance the effectiveness of bug fixing. As such, \approach optimizes the capability of LLMs to produce correct patches that accurately fix the given bugs. In this paper, we employ the potent dialogue-based ChatGPT model, and concentrate on resolving single-line bugs written in the Java programming language, a task frequently examined in prior studies \cite{zhong2022standupnpr}. Typically, a single-turn conversation involves taking the system and user messages as input and returning an assistant message generated by ChatGPT as output. The system message plays a crucial role in defining the behavior of the assistant, while the user message serves as a means to convey requests or comments for the assistant to respond to. We will provide detailed descriptions of the specific prompting used for each ChatGPT agent in subsequent subsections.

\subsection{Tester}

\begin{figure}[h]
  \centering
  \includegraphics[width=0.48\textwidth]{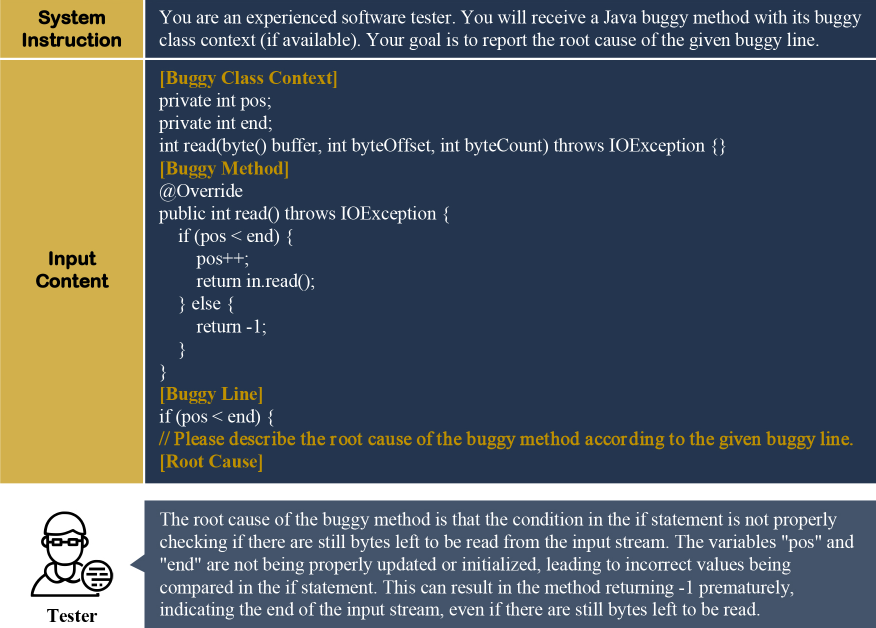}
  \caption{A prompting example of the tester's behavior during the bug reporting stage.}
  \label{reporting}
\end{figure}

Figure~\ref{reporting} illustrates the prompt and the corresponding output when simulating the tester's behavior during the bug reporting stage. The \textbf{\texttt{System Instruction}} (i.e., system message) specifies the persona adopted by the ChatGPT agent in its responses. The main objective of the tester is to report the root cause of the given buggy method based on its fault location (i.e., buggy line). In order to obtain a highly relevant response, the \textbf{\texttt{Input Content}} (i.e., user message) provides crucial details and context to the ChatGPT agent. Moreover, we use delimiters (highlighted in bold {\color[RGB]{183,140,4}\textbf{orange}}) to clearly indicate distinct parts of the \textbf{\texttt{Input Content}}. As shown in Fig.\ref{reporting}, following the provided prompt, the tester produces a bug report (i.e., assistant message) that describes the nature of the bug and its impact on the given buggy method. These information will then be utilized to aid the developer in resolving the bug.

\subsection{Developer}

In the context of \approach, the responsibilities of the developer encompass four main aspects: 1) Summarizing bug-fixing patterns by analyzing similar demonstrations retrieved from the historical code corpus. 2) Elucidating the given buggy code line-by-line using rubber duck debugging techniques. 3) Generating the initial candidate patch using information from the bug report and guidance obtained from the bug diagnosis process. 4) Refining the candidate patch by incorporating feedback received from the reviewer.

\begin{figure*}[t]
  \centering
  \includegraphics[width=\textwidth]{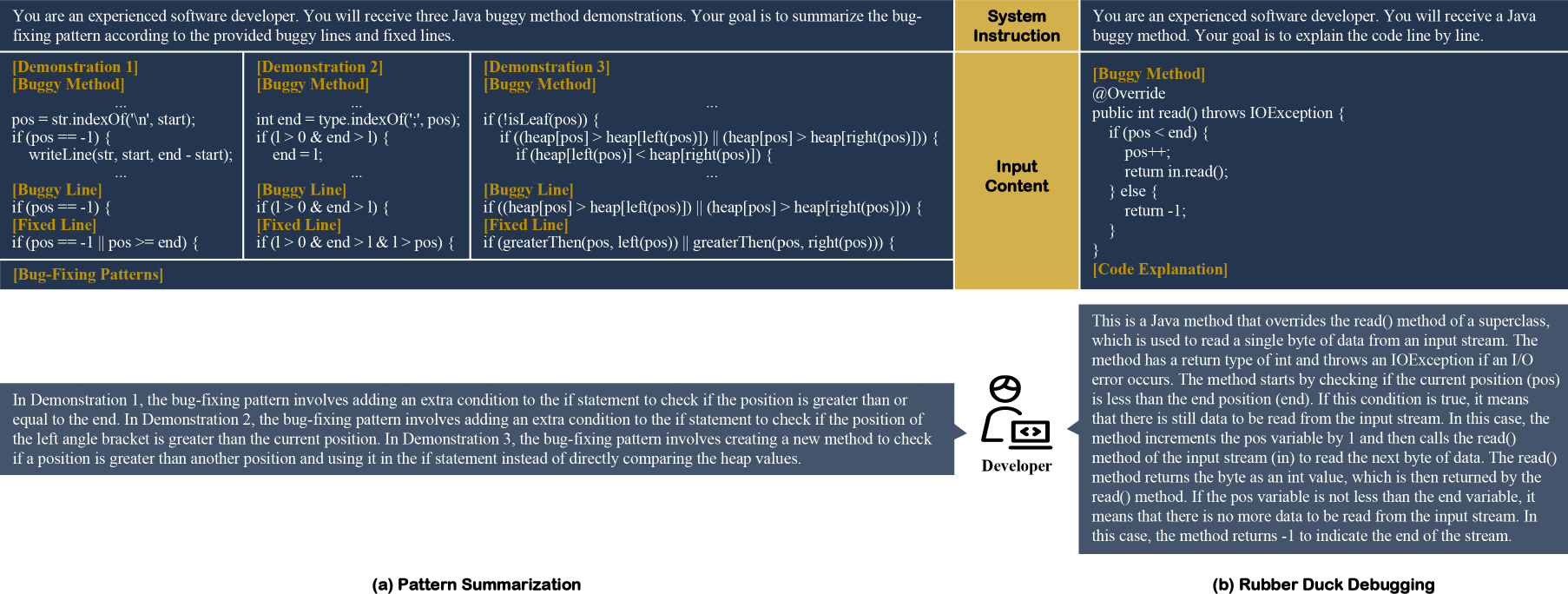}
  \caption{A prompting example of the developer's behavior during the bug diagnosis stage.}
  \label{diagnosis}
\end{figure*}

\subsubsection{Bug Diagnosis}
\label{bugdiag}

Figure~\ref{diagnosis} illustrates the example prompt along with the corresponding output achieved during the simulation of the developer's behavior in the bug diagnosis stage. The primary objective of the developer is to initially summarize the bug-fixing patterns by analyzing the paired buggy and fixed lines. Subsequently, the developer aims to provide a detailed, line-by-line explanation of the given buggy code in natural language.

\textbf{(a) Pattern Summarization.} As depicted in the left part of Fig.\ref{diagnosis}, the initial step of this process involves retrieving similar programs from a historical code corpus based on both the given buggy method and buggy line. To achieve this, \approach utilizes the BM25 score \cite{robertson2009probabilistic} as the retrieval metric, a probabilistic model-based scoring algorithm widely employed in previous studies \cite{wei2020retrieve,li2021editsum}. BM25 functions as a bag-of-words retrieval method and estimates lexical-level similarity between two sentences. Higher BM25 scores indicate greater similarity between sentences. Specifically, \approach selects the top-3 demonstrations, including buggy methods and paired buggy and fixed lines, as the retrieved results from the historical corpus. Based on the selected demonstrations, the developer summarizes the bug-fixing patterns, providing insights into the root causes and resolutions of similar issues.

\textbf{(b) Rubber Duck Debugging.} As shown in the right part of Fig.\ref{diagnosis}, this debugging process emulates the common practice among human programmers, where they explain the code line-by-line in natural language, as if talking to a rubber duck \cite{spinellis2000pragmatic}. Consequently, the developer provides the code explanation by describing the code implementation, which enhances the debugging efficiency without the need for additional guidance, such as unit tests.

\subsubsection{Patch Generation}

\begin{figure}[htbp]
  \centering
  \includegraphics[width=0.48\textwidth]{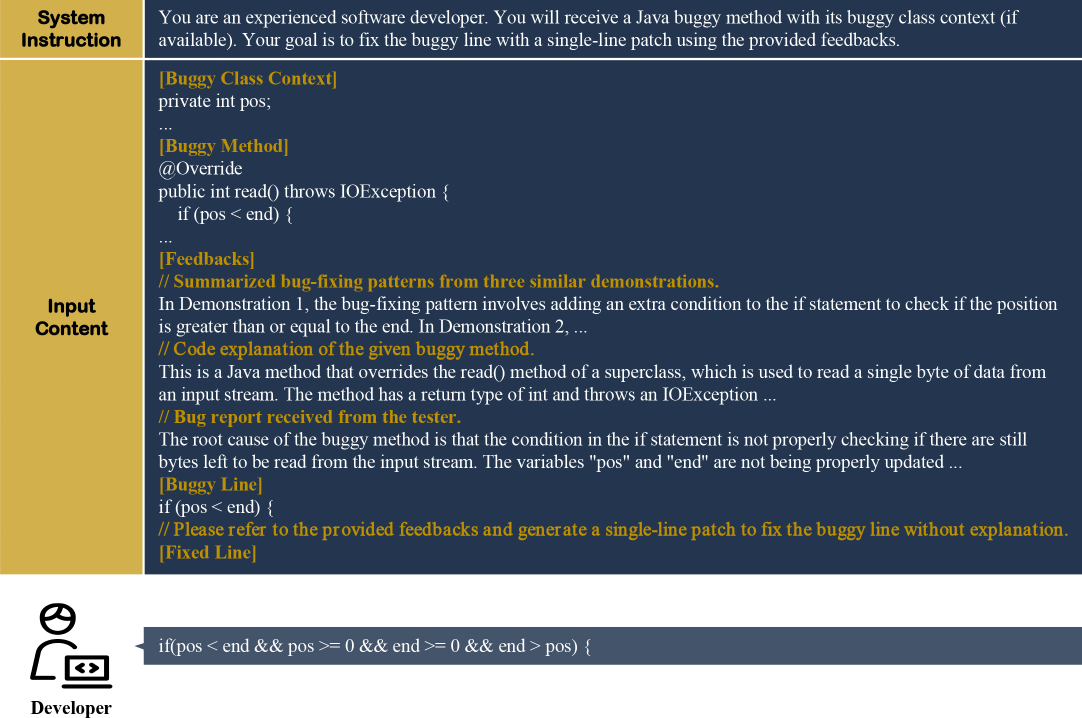}
  \caption{A prompting example of the developer's behavior during the patch generation stage.}
  \label{generation}
\end{figure}

As illustrated in Fig.\ref{generation}, the developer's objective (stated in the \textbf{\texttt{System Instruction}}) at this stage is to fix the buggy line by producing a single-line patch utilizing the feedbacks enumerated in the \textbf{\texttt{Input Content}}. In previous studies, LLMs are prompted directly to generate patches based on the provided buggy code. Consequently, challenges arise in the accuracy of patch generation. In light of this, \approach provides the developer with a guided process for bug reporting and diagnosis, enabling the developer to generate the initial candidate patch by incorporating information from the bug report, bug-fixing patterns, and code explanation as prompt. Afterward, the generated candidate patch undergoes further verification by the reviewer.

\subsection{Reviewer}

\begin{figure*}[t]
  \centering
  \includegraphics[width=\textwidth]{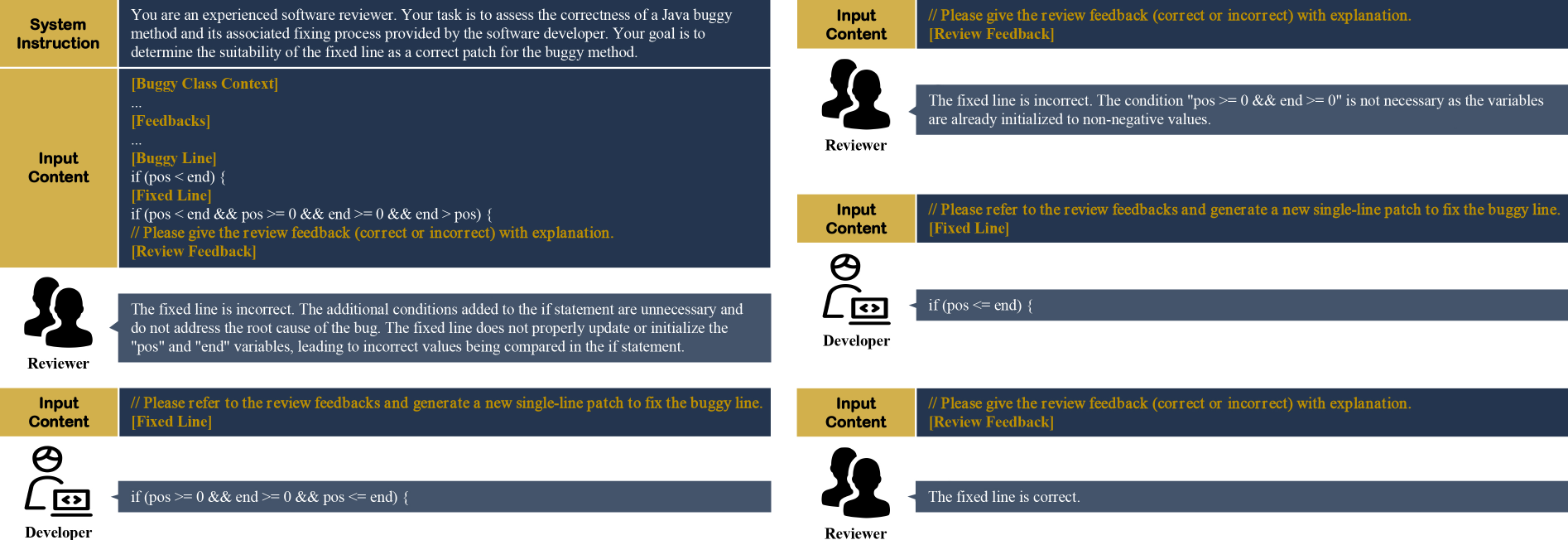}
  \caption{A prompting example of the interactive behavior between reviewer and developer during the patch verification stage.}
  \label{verification}
\end{figure*}

As described in the \textbf{\texttt{System Instruction}} in Fig.\ref{verification}, the reviewer acquires the buggy method and the associated fixing process provided by the developer. At this stage, the goal of the reviewer is to infer the correctness of the candidate patch and deliver feedback messages to the developer for facilitating subsequent interactive steps. Figure~\ref{verification} outlines an exemplary interactive process between the reviewer and developer during the patch verification stage. If the reviewer determines the fixed line as an incorrect patch for the buggy method, the developer is required to generate a new single-line patch while taking into account the review feedbacks. The interactive process terminates either when the reviewer confirms the correctness of the fixed line or when the maximum allowed number of verification turns is reached.

\section{Experimental Setup}
\label{exp}

In this section, we outline the comprehensive setups of our study, including research questions, experimental benchmarks, evaluation metrics, comparison baselines, and implementation details.

\subsection{Research Question}

To assess the effectiveness of \approach, we conduct extensive experiments to answer three research questions (RQs): 
\begin{itemize}
  \item \textbf{RQ1: How does \approach compare against the state-of-the-art baselines?} The objective of this RQ is to assess the superior effectiveness of \approach compared to state-of-the-art baselines in the context of bug fixing. To achieve this, we conduct a comprehensive comparison of \approach against 11 LLMs using a widely adopted bug-fixing benchmark.
  \item \textbf{RQ2: How does each component impact the performance of \approach?} The proposed framework comprises three essential components: the tester, the developer, and the reviewer. The tester is responsible for bug reporting, the developer handles bug diagnosis and patch generation, and the reviewer is in charge of patch verification. In this RQ, we aim to analyze the contributions of each designed component by conducting an ablation study.
  \item \textbf{RQ3: What is the generalizability of \approach for additional benchmarks?} This RQ aims to evaluate the generalizability of \approach across various benchmarks that are commonly used in the program repair task. To achieve this goal, we incorporate four additional benchmarks to enrich the evaluation diversity and conduct a comparative analysis between \approach and eight traditional approaches. 
\end{itemize}

\subsection{Benchmark}

To evaluate \approach, we employ the widely used benchmark of BFP \cite{tufano2019empirical}, which encompasses a substantial volume of real-world bug-fixing commits extracted from the GitHub repositories. Each instance in the BFP benchmark consists of both the buggy and fixed versions of a Java method. In order to provide the necessary global context of the buggy method for better comprehension of the buggy code, we utilize the static analysis tool Spoon \cite{pawlak2016spoon} to parse each Java method into an abstract syntax tree for context extraction. During this step, we filter out instances that could not be successfully parsed by Spoon and also exclude instances exceeding 150 tokens in length, considering the max token limits of the GPT family of models. Next, we perform a meticulous split of the BFP dataset into training, validation, and testing sets in an 8:1:1 ratio while ensuring data leakage prevention. To be specific, instances originating from the same GitHub repository cannot allowed to be present in different sets (e.g., one in training and the other in testing). In total, we collect 18226 bug-fixing pairs in the training set, 2292 in the validation set, and 2292 in the testing set. Notably, each instance represents a single-line bug that can be fixed using a single-line patch within a Java method. We measure the performance of \approach and selected baselines on the testing set. Additionally, we conduct the pattern summarization process, as described in Section~\ref{bugdiag}, by retrieving similar instances from the training set.

\subsection{Metric}

In order to quantitatively compare the performance of \approach with the baselines, we choose the following three evaluation metrics:
\begin{itemize}
  \item \textbf{Fix@k.} This paper employs the Fix@k metric to evaluate the model performance on the testing set. To be specific, given a Java method with a single-line bug, the corresponding LLM is permitted to generate $k$ candidate patches. The bug is considered resolved if any of the generated patches match the human-written ground truth. Fix@k denotes the percentage of successfully fixed bugs in the entire testing set. In this paper, $k$ is set to 1.
  \item \textbf{BLEU.} The BLEU score \cite{papineni2002bleu} calculates similarity by measuring the n-gram precision of a candidate patch with respect to the human-written ground truth, while also penalizing overly short length. A higher BLEU score indicates a closer resemblance between the candidate patch and the ground truth. In this paper, we report the BLEU-4 score.
  \item \textbf{Levenshtein Distance.} This evaluation metric computes the absolute token-based edit distance between the candidate patch and the ground truth (i.e., the minimum number of operations needed to transform the candidate patch into the ground truth). A lower Levenshtein distance indicates a closer match between the candidate patch and the ground truth. This metric provides valuable insights into the usefulness of incorrect predictions for developers.
\end{itemize}

\subsection{Baseline}

This paper centers on addressing the bug-fixing task using LLMs. Therefore, we compare \approach against state-of-the-art LLMs as baselines. Table~\ref{benchmark} presents the 11 LLMs evaluated in this paper. The selection process for the LLMs is based on the following criteria:
\begin{itemize}
  \item \textbf{Popularity.} Initially, we consider the list of popular models hosted on the Hugging Face website, which is an open-source platform for hosting and deploying large models. Among these models, we choose those that are trained on a substantial code corpus. Additionally, we include closed-source models (i.e., Codex and ChatGPT) as they have demonstrated impressive performance on code-related tasks. 
  \item \textbf{Diversity.} To ensure a diverse set of models, we select models with varying sizes of parameters and from different organizations (listed in Column \textbf{\texttt{Size}} and \textbf{\texttt{Institute}}).
  \item \textbf{Accessibility.} The LLMs evaluated in this paper are publicly accessible either through checkpoints (e.g., CodeGen) or APIs (e.g., Codex). As a result, we have excluded the closed-source models such as AlphaCode \cite{li2022competition}.
\end{itemize}

\begin{table}[ht]
    \centering
    \caption{Overview of the evaluation LLMs.}
    \label{benchmark}
    \resizebox{0.48\textwidth}{!}{
    \begin{tabular}{lrcc}
        \toprule
        \multicolumn{1}{c}{\textbf{Model}} & \multicolumn{1}{c}{\textbf{Size}} & \textbf{Institute} & \textbf{Pre-Training Code Corpus} \\
        \midrule
        CodeParrot \cite{tunstall2022natural} & 110M & Hugging Face & CodeParrot \\
        CodeGPT \cite{lu2021codexglue} & 124M & Microsoft & CodeSearchNet \\
        GPT-Neo \cite{sid2021gptneo} & 2.7B & EleutherAI & Pile \\
        PolyCoder \cite{xu2022systematic} & 2.7B & CMU & GitHub \\
        CodeGen \cite{nijkamp2023codegen} & 6.1B & Salesforce & Pile \& BigQuery \& BigPython \\
        InCoder \cite{fried2023incoder} & 6.7B & Facebook & StackOverFlow \& GitHub \& GitLab \\
        FLAN-T5 \cite{chung2022scaling} & 11B & Google & Muffin \\
        LLaMA \cite{touvron2023llama} & 13B & Meta & BigQuery \\
        GPT-NeoX \cite{black2022gptneox} & 20B & EleutherAI & Pile \\
        Codex \cite{chen2021evaluating} & 175B & OpenAI & - \\
        ChatGPT \cite{openai2023chatgpt} & - & OpenAI & - \\
        \bottomrule
    \end{tabular}}
\end{table}

\subsection{Implementation}
\label{implementation}

We implement the main logic of \approach in Python by invoking ChatGPT through its API. As our base model, we employ the stable \texttt{gpt-3.5-turbo-0301} version of of the ChatGPT family to minimize the risk of unexpected model changes affecting the results. Following the best-practice guide \cite{shieh2023best}, we design prompts and manually examine a few alternative approaches with selected buggy code using the Web-version of ChatGPT. The configurations for each ChatGPT agent are detailed as follows:
\begin{itemize}
  \item \textbf{Tester.} To simulate the tester's behavior, the prompting format is displayed in Fig.\ref{reporting}. The maximum generated length of the bug report is restricted to 200 tokens.
  \item \textbf{Developer.} For simulating the developer's behavior, two prompting formats are shown in Fig.\ref{diagnosis} and Fig.\ref{generation}. The maximum generated length of the bug-fixing pattern and code explanation is limited to 500 tokens, while the length of the generated candidate patch is capped at 150 tokens.
  \item \textbf{Reviewer.} To simulate the behavior of the reviewer, the prompting format is presented in Fig.\ref{verification}. The maximum generated length of the review feedback is constrained to 200 tokens. Furthermore, we set the number of interaction turns between the reviewer and developer to be 3, which is based on empirical suggestions by Chen et al. \cite{chen2023teaching}.
\end{itemize}
In all experiments, we utilize greedy decoding to generate feedback messages (i.e., bug reports, bug-fixing patterns, code explanations, and review feedbacks) and candidate patches, namely, \approach generates the top-1 chat completion choice for each input message. Specifically, we employ a sampling temperature of 0 to enhance the stability of the LLM's output. Furthermore, we conduct experiments under the zero-shot setting, where task examples are not provided, aiming to demonstrate the superiority of our proposed framework.

\section{Results and Analysis}
\label{res}

\subsection{Answering RQ1}

To answer this question, we conduct a comprehensive comparison of \approach with 11 state-of-the-art baselines on the BFP benchmark. Each baseline is implemented by either reusing the official checkpoint or accessing the inference API. Consistent with previous studies, we prompt the baselines with solely the buggy method of the testing set. To ensure the fairness of comparison, we employ the same hyper-parameters of sampling as described in Section~\ref{implementation}.

\subsubsection{Experimental Metric Evaluation}

Table~\ref{rq1a} presents the bug-fixing performance of different models in terms of the three evaluation metrics. The best result for each metric is highlighted in bold. Our experiments yield the following three-fold findings:
\begin{enumerate}
	\item \textbf{\approach exhibits superior performance compared to all the baselines on the BFP benchmark.} To be specific, \approach surpasses the best baseline ChatGPT by 10.9\% in terms of producing correct patches. These improvements highlight the prowess of \approach in the bug fixing task as \textbf{Fix@1} is a strict metric. Additionally, for the \textbf{BLEU-4} and \textbf{Levenshtein Distance} metrics, \approach achieves scores of 72.31 and 21.44, respectively, on the BFP benchmark, showcasing improvements of 22.8\% and 35.6\% over ChatGPT.
	\item \textbf{Simulating programmer behavior proves to be advantageous for bug fixing.} \approach does not alter the parameters of ChatGPT; instead, it explicitly instructs ChatGPT to mimic the behavior of programmers engaged in the bug management process. The substantial improvements observed over ChatGPT indicate that \approach effectively endows ChatGPT with collaborative problem-solving abilities, thereby enhancing its bug-fixing capabilities.
	\item \textbf{Enhancing the performance of LLMs relies on having more parameters and well-designed prompts.} In particular, an increase in parameters often leads to improved performance, as exemplified by Codex-175B surpassing GPT-NeoX-20B, while GPT-NeoX-20B performs better than LLaMA-13B. Notably, LLMs struggle to achieve satisfactory performance under the zero-shot setting, due to the lack of task examples and their inability to comprehend how to solve the given problem. However, this limitation can be effectively addressed by incorporating crucial information in the prompts. \approach significantly outperforms the baseline LLMs after adopting this approach. This finding validates our motivation to decompose the bug fixing task into subtasks using well-designed prompts, as it substantially enhances the performance of LLMs in this context.
\end{enumerate}

\begin{table}[ht]
    \centering
    \caption{Comparison of \approach against the baselines.}
    \label{rq1a}
    \resizebox{0.48\textwidth}{!}{
    \begin{tabular}{lccc}
        \toprule
        \multicolumn{1}{c}{\textbf{Model}} & \textbf{Fix@1 (\%) $\uparrow$} & \textbf{BLEU-4 $\uparrow$} & \textbf{Levenshtein Distance $\downarrow$} \\
        \midrule
        CodeParrot & 2.75 & 16.80 & 70.34 \\
        CodeGPT & 2.79 & 14.19 & 73.69 \\
        GPT-Neo & 2.88 & 29.90 & 57.37 \\
        PolyCoder & 1.88 & 5.59 & 80.18 \\
        CodeGen & 2.97 & 20.55 & 70.13 \\
        InCoder & 2.01 & 23.00 & 64.20 \\
        FLAN-T5 & 2.05 & 12.37 & 71.46 \\
        LLaMA & 2.53 & 19.46 & 71.39 \\
        GPT-NeoX & 4.45 & 54.61 & 36.33 \\
        Codex & 9.77 & 55.49 & 34.27 \\
        ChatGPT & 10.95 & 58.89 & 33.28 \\
        \midrule
        \approach & \textbf{21.86} & \textbf{72.31} & \textbf{21.44} \\
        \bottomrule
    \end{tabular}}
\end{table}

\begin{figure*}[b]
  \centering
  \includegraphics[width=\textwidth]{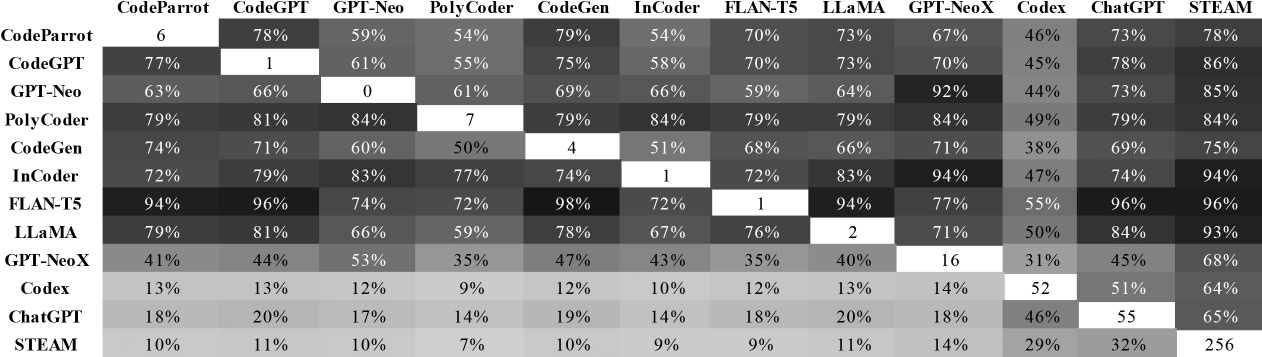}
  \caption{The overlapping rates and unique patch numbers of the evaluated models.}
  \label{rq1b}
\end{figure*}

\subsubsection{Overlapping Phenomenon Evaluation}

As illustrated in Fig.\ref{rq1b}, each row represents the overlapping ratio of correct patches generated by one model and the others, while the diagonal indicates the number of unique correct patches generated by each model on the BFP benchmark. As the overlapping rate increases, the color of the rectangle darkens. For instance, \approach (row 12) generates correct patches that overlap with 32\% of the patches generated by CodeX (column 11). Additionally, there are 256 bugs (row 12, column 12) that can only be fixed by \approach. The results in Fig.\ref{rq1b} indicate that models with better fixing performance tend to have higher overlapping patching rates with other models. Regarding the evaluation results in Table~\ref{rq1a}, we can find that \approach, ChatGPT, and Codex are the top three models. In comparison, the overlapping rates of other models with these three are notably higher. This phenomenon could be attributed to the adoption of similar network architectures and inference paradigms among DL-based approaches. Furthermore, \approach generates a larger number of unique correct patches compared to other baseline models.

\begin{tcolorbox}[size=title]
    \textbf{Answer to RQ1}: In conclusion, the proposed framework exhibits substantial superiority over the baselines in the three evaluation metrics, highlighting the effectiveness of \approach in the bug fixing task. Furthermore, our observations indicate that \approach has the ability to generate a greater number of unique and correct patches compared to the baselines.
\end{tcolorbox}

\subsection{Answering RQ2}

To answer this question, we conduct ablation experiments to evaluate the impact of different components in the \approach design. To ensure the fairness of comparison, we maintain consistency in the experimental settings with those detailed in Section~\ref{implementation}.

\subsubsection{Ablation Study}

\begin{figure*}[tb]
  \centering
  \includegraphics[width=\textwidth]{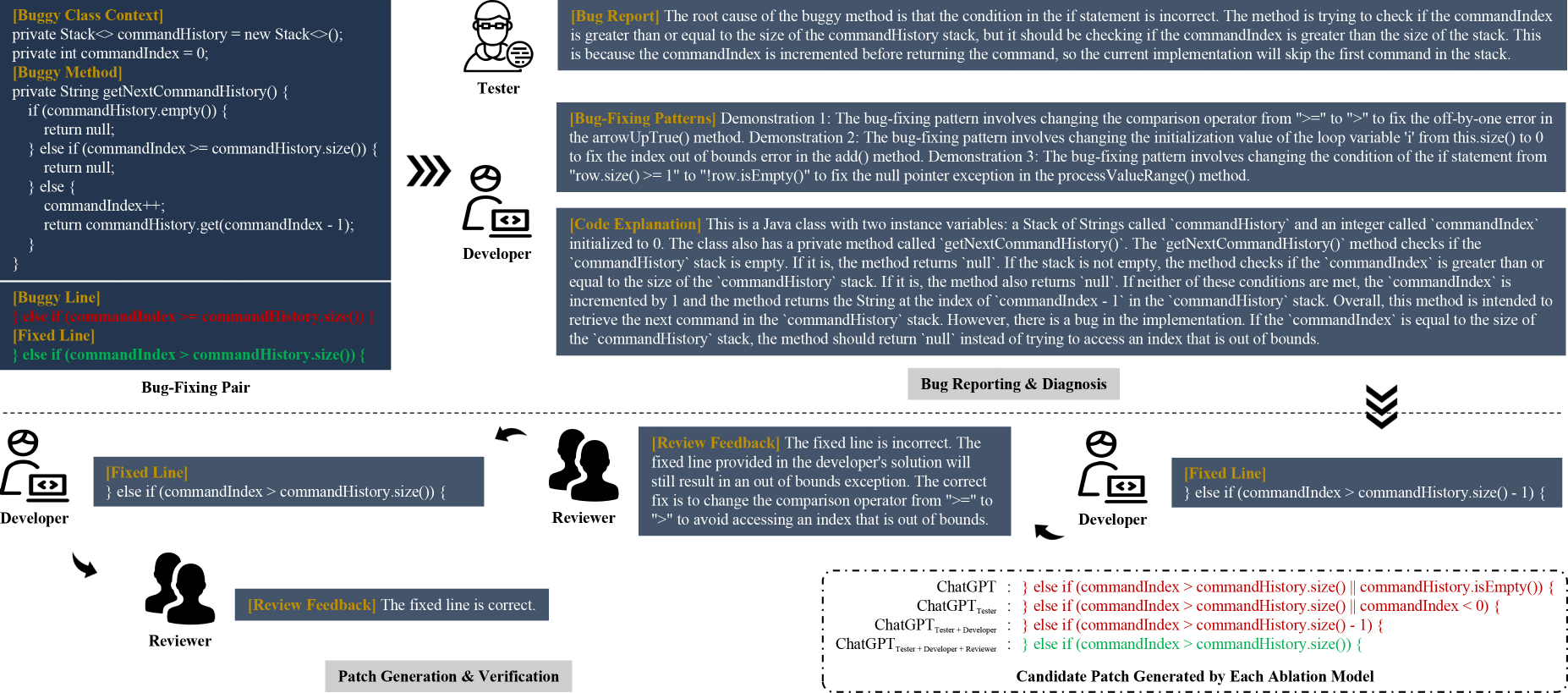}
  \caption{An example from BFP only fixed by \approach.}
  \label{rq2a2}
\end{figure*}

Table~\ref{rq2a1} presents the evaluation results with each row representing one ablated model. The symbols \faCheck\ and \faTimes\ denote the addition and removal of corresponding components, respectively. The best result for each metric is marked in bold. To illustrate how each component contributes to the bug-fixing performance, we begin with the basic LLM---ChatGPT, which employs only the buggy method as a prompt to generate candidate patches. When augmenting the \texttt{Tester} component, the ChatGPT agent initially proceeds bug reporting to outline the underlying cause of the buggy code and then generates candidate patches based on the information available in the bug report. With the addition of the \texttt{Developer} component, the ChatGPT agent can generate candidate patches under the guidance of the bug report, bug-fixing patterns, and code explanations. Furthermore, the ChatGPT agent gains interactive abilities to refine the candidate patches based on review feedbacks after incorporating the \texttt{Reviewer} component. Significantly, all components are crucial for the optimal performance of \approach. Specifically, with the incorporation of the \texttt{Tester} component, the performance of ChatGPT is respectively improved by 23.6\%, 5.5\% and 4.4\% in terms of the three evaluation metrics. This underscores the significance of providing essential bug-related information in the bug-fixing process. The \texttt{Developer} component further enhances \textbf{Fix@1} by 27.1\%, \textbf{BLEU-4} by 4.4\%, and \textbf{Levenshtein Distance} by 13.6\%, demonstrating the effectiveness of the self-guided diagnosis process in improving bug-fixing efficiency. Moreover, the addition of the \texttt{Reviewer} component leads to continuous improvements in bug-fixing performance, with enhancements of 27.2\% in \textbf{Fix@1}, 11.5\% in \textbf{BLEU-4}, and 22.0\% in \textbf{Levenshtein Distance}. This highlights the importance of interaction and collaboration during the resolution of software bugs. Figure~\ref{rq2a2} illustrates a bug-fixing example from the BFP benchmark, wherein only \approach correctly patches the bug. In this case, the root cause of the bug lies in an incorrect condition in the if statement of the buggy line. Consequently, a correct patch must verify whether the commandIndex is greater than the size of the stack. As indicated in the lower right corner of Fig.\ref{rq2a2}, we can observe that the ablated model without the \texttt{Reviewer} component (i.e., ChatGPT$_\mathtt{Tester+Developer}$) generates an incorrect patch that is semantically equivalent to the buggy line, resulting in an out-of-bounds exception even after the patch is applied. However, with the collaboration of the \texttt{Reviewer} component, \approach (i.e., ChatGPT$_\mathtt{Tester+Developer+Reviewer}$) successfully generates the correct patch that is identical to the ground truth.

\begin{table}[ht]
    \centering
    \caption{Ablation study for \approach.}
    \label{rq2a1}
    \resizebox{0.48\textwidth}{!}{
    \begin{tabular}{ccccccc}
        \toprule
        \multirow{2}{*}{\textbf{Model}} & \multicolumn{3}{c}{\textbf{Component}} & \multirow{2}{*}{\textbf{Fix@1 (\%) $\uparrow$}} & \multirow{2}{*}{\textbf{BLEU-4 $\uparrow$}} & \multirow{2}{*}{\textbf{Levenshtein Distance $\downarrow$}} \\
        \cmidrule[0.5pt](rl){2-4}
        & \texttt{Tester} & \texttt{Developer} & \texttt{Reviewer} & & & \\
        \midrule
        \multirow{4}{*}{ChatGPT} & \faTimes & \faTimes & \faTimes & 10.95 & 58.89 & 33.28 \\
        & \faCheck & \faTimes & \faTimes & 13.53 & 62.10 & 31.82 \\
        & \faCheck & \faCheck & \faTimes & 17.19 & 64.83 & 27.48 \\
        & \faCheck & \faCheck & \faCheck & \textbf{21.86} & \textbf{72.31} & \textbf{21.44} \\
        \bottomrule
    \end{tabular}}
\end{table}

\subsubsection{The Impact of Interaction Turns}

In order to assess the impact of reviewer-developer interaction, we control the number of interaction turns during this experiment, and the corresponding results are presented in Table~\ref{rq2b}. When the number of interaction turns is set to zero, it indicates a complete absence of interaction between the programmers involved. This means that the candidate patch generated by the developer does not receive any form of feedback from the reviewer, resulting in the same outcome as the ablated model ChatGPT$_\mathtt{Tester+Developer}$. It is worth noting that the most significant improvement arises from the first interaction turn. Specifically, a single interaction turn with the \texttt{Reviewer} component leads to an approximately 8\% enhancement in terms of \textbf{Fix@1} over the ChatGPT$_\mathtt{Tester+Developer}$ model. As the number of interaction turns continues to increase beyond the initial round, the improvements tend to diminish; however, a consistent enhancement is still observed. This suggests that the ability to fix more complex bugs is still achieved through additional interactions.

\begin{table}[ht]
    \centering
    \caption{The effect of interaction turns on bug fixing.}
    \label{rq2b}
    \resizebox{0.48\textwidth}{!}{
    \begin{tabular}{cccc}
        \toprule
        \textbf{\# of Turns} & \textbf{Fix@1 (\%) $\uparrow$} & \textbf{BLEU-4 $\uparrow$} & \textbf{Levenshtein Distance $\downarrow$} \\
        \midrule
        0 & 10.95 & 58.89 & 33.28 \\
        1 & 18.80 & 66.45 & 26.71 \\
        2 & 20.77 & 69.27 & 23.94 \\
        3 & \textbf{21.86} & \textbf{72.31} & \textbf{21.44} \\
        \bottomrule
    \end{tabular}}
\end{table}

\begin{tcolorbox}[size=title]
    \textbf{Answer to RQ2}: To sum up, all components of \approach can significantly contribute to performance improvements. Regarding Fix@1, the addition of the tester enhances ChatGPT by 23.6\%. Furthermore, the introduction of the developer leads to additional improvements of 27.1\%, while the incorporation of the reviewer results in continuous enhancements of 27.2\%.
\end{tcolorbox}

\subsection{Answering RQ3}

To answer this question, we conduct a thorough evaluation of the generalizability of \approach across four widely-used benchmarks in the automated program repair (APR) domain, specifically, Bugs.jar \cite{saha2018bugsjar}, Defects4J \cite{just2014defects4j}, Bears \cite{delfim2019bears}, and QuixBugs \cite{lin2017quixbugs}. We compare \approach against eight baselines encompassing traditional, neural-based, and pre-trained language model-based APR approaches. For traditional APR, we employ state-of-the-art template-based baseline TBar \cite{liu2019tbar} with perfect fault localization configuration. As for neural-based APR, we select five recently published approaches, namely Tufano \cite{tufano2019empirical}, CoCoNut \cite{lutellier2020coconut}, {\sc SequenceR\xspace} \cite{chen2021sequencer}, Recoder \cite{zhu2021syntax}, and {\sc RepeatNPR\xspace} \cite{zhang2023neural}. Additionally, we include two pre-trained models, CodeBERT \cite{feng2020codebert} and CodeT5 \cite{wang2021codet5}. In line with previous studies \cite{zhong2022standupnpr}, we use the same training strategy and hyper-parameter settings to ensure a fair comparison. In this experiment, we adopt an objective metric \textit{exact match} to assess the correctness of each generated candidate patch. This evaluation method helps avoid human bias and reduces the need for manual effort in the assessment of model performance on the four APR benchmarks.

\begin{table}[t]
\centering
    \caption{Comparison of \approach on four APR benchmarks against eight baselines.}
\label{rq3a}
\resizebox{0.48\textwidth}{!}{
    \begin{tabular}{lcccc}
    \toprule
    \multicolumn{1}{c}{\multirow{2}{*}{\textbf{Model}}} & \textbf{Bugs.jar} & \textbf{Defects4J} & \textbf{Bears} & \textbf{QuixBugs} \\
    \cmidrule[0.5pt](rl){2-5}
    & {1000 bugs} & {260 bugs} & {119 bugs} & {32 bugs} \\
    \midrule
    {TBar} & - & 43 & - & - \\
    \midrule
    {Tufano} & 56 & 18 & 8 & 7 \\
    {Recoder} & 61 & 33 & 1 & 10 \\
    {CoCoNut} & 66 & 37 & 16 & 13 \\
    {{\sc {SequenceR}}\xspace} & 99 & 38 & 14 & 15 \\
    {{\sc {RepeatNPR}}\xspace} & 168 & 44 & 24 & 15 \\
    \midrule
    {CodeBERT} & 111 & 29 & 12 & 7 \\
    {CodeT5} & 150 & 36 & 16 & 14 \\
    \midrule
    \approach & \textbf{180} & \textbf{65} & \textbf{25} & \textbf{26} \\
    \bottomrule
    \end{tabular}
}
\end{table}

Table~\ref{rq3a} reports the results of the evaluation, displaying the number of correct patches that are identical to the human-written ones for both \approach and the eight baseline models. The best performance for each benchmark is indicated in bold. As observed in Table~\ref{rq3a}, \approach exhibits a substantial performance advantage over the compared baselines across all four benchmarks. Specifically, \approach outperforms the best baseline model {\sc RepeatNPR\xspace} by 7.1\% in the Bugs.jar benchmark, 47.7\% in the Defects4J benchmark, 4.2\% in the Bears benchmark, and an impressive 73.3\% in the QuixBugs benchmark. As described in Section~\ref{implementation}, \approach generates the top-1 candidate patch for each bug, whereas the selected baselines typically evaluate a larger number of candidates in prior studies. In this experiment, we report the number of correct patches within the top-10 candidates generated by each baseline, aligning with recent findings \cite{noller2022trust} that most developers are only willing to review up to 10 patches. We further investigate the number of correct patches generated by each baseline on the Bugs.jar benchmark under different candidate numbers, considering various candidate numbers (1, 5, and 10). As depicted in Table~\ref{rq3b}, a large candidate set clearly increases the likelihood of containing the correct patch. Notably, with the increasing candidate numbers, all baselines consistently demonstrated improved performance gains. Nevertheless, it is noteworthy that even with 10 candidates, applying \approach to greedy decoding continues to outperform the baselines.

\begin{table*}[t]
\centering
    \caption{The number of correct patches generated by each baseline on Bugs.jar under different candidate numbers.}
\label{rq3b}
    \begin{tabular}{ccccccccc}
    \toprule
    \textbf{\# of Candidates} & Tufano & Recoder & CoCoNut & {{\sc {SequenceR}}\xspace} & {{\sc {RepeatNPR}}\xspace} & CodeBERT & CodeT5 & \approach \\
    \midrule
    1 & 20 & 26 & 21 & 32 & 67 & 32 & 52 & 180 \\
    5 & 40 & 55 & 39 & 78 & 133 & 86 & 119 & - \\
    10 & 56 & 61 & 66 & 99 & 168 & 111 & 150 & - \\
    \bottomrule
    \end{tabular}
\end{table*}

\begin{tcolorbox}[size=title]
    \textbf{Answer to RQ3}: In contrast to template-based or neural-based approaches, \approach does not rely on fine-tuning with specific bug-fixing datasets, making it less susceptible to generalizability issues. As a result, \approach outperforms traditional approaches across various APR benchmarks. Looking ahead, \approach has the potential to be seamlessly integrated with more robust LLMs in a plug-and-play manner.
\end{tcolorbox}

\section{Threats to Validity}
\label{thr}

In this section, we illustrate the main threats to the validity of our approach, which are listed as follows:
\begin{itemize}
	\item \textbf{External threat}: The primary threats to external validity in this paper revolve around the quality of the selected experimental subjects and the generalizability of \approach. It is uncertain whether the improvements achieved by \approach will apply to other bug-fixing benchmarks. To address this concern, we have adopted the mainstream benchmark BFP, consistent with prior studies\cite{tufano2019empirical,tang2021grammar,wang2021codet5,zhong2022standupnpr}, and supplemented the evaluation with four additional APR benchmarks to enhance the evaluation diversity. Moreover, \approach specifically targets single-line Java bugs in this study. However, it is important to note that the designed components in \approach are language-agnostic and can be effectively applied to other programming languages.
	\item \textbf{Internal threat}: LLMs are known to be sensitive to prompts and hyper-parameters, particularly the number of examples and natural language instructions, which can significantly impact their performance. To alleviate this threat, we employ the same prompts and hyper-parameters for \approach and baselines. We refrain from experimental tuning of the prompt design and hyper-parameters, and set them empirically. Thus, we acknowledge that further improvement may be attainable through additional tuning.
	\item \textbf{Construct threat}: In this paper, the experimental metric used for model evaluation is referred to as the construct threat. Specifically, the Fix@1 metric is adopted to assess the correctness of the generated candidate patches. Although this metric does not reflect human judgment, it serves as a strict and objective measure that allows for quick and quantitative evaluation of the model's performance. In the future, we plan to conduct additional human evaluations to further validate the models.
\end{itemize}

\section{Conclusion and Future Work}
\label{con}

In this paper, we present a stage-wise framework aimed at enhancing the bug-fixing capabilities of LLMs in a interactive and collaborative manner. We explore the potential of ChatGPT in the bug-fixing task, simulating the behavior of programmers involved in bug management. Specifically, we decompose the bug fixing task into four distinct stages and employ three ChatGPT agents, each responsible for specific stages. These agents generate correct patches to fix the bugs collaboratively using prompts. Extensive experiments are conducted to demonstrate the effectiveness and generalizability of \approach. We firmly believe that aligning the collaborative problem-solving abilities of programmers with LLMs represents a pivotal stride toward intelligent software engineering.

\bibliographystyle{IEEEtran}
\bibliography{icse}

\end{document}